\documentclass[12pt]{article}
\pdfoutput=1
\usepackage{hyperref}
\usepackage{graphicx}
\usepackage{subcaption}
\graphicspath{{./picture/}}
\usepackage{empheq}
\usepackage{amssymb}
\usepackage{mathtools}
\usepackage{commath}
\usepackage{verbatim}
\usepackage{xcolor}
\usepackage{soul}
\usepackage{booktabs}
\usepackage{amsmath}
\usepackage{hhline}
\usepackage[scale={0.75,0.8}]{geometry}
\usepackage{url}
\usepackage{caption}
\usepackage[numbers, sort&compress]{natbib}
\usepackage{epsfig}
\usepackage{lscape}
\usepackage{mathrsfs}
\usepackage{nicefrac} 
\usepackage{upgreek}
\usepackage{bbm}
\usepackage{psfrag}
\usepackage{hyperref}
\usepackage{hyperref}

\renewcommand{\Re}{\text{Re}}
\renewcommand{\Im}{{\text Im}}

\newcommand{\iu}{{\rm i}}
\newcommand{\Mu}{\mu}
\newcommand{\M}{\bar{M}}
\newcommand\ii{{\rm i}}		
\usepackage{slashed}				

\begin{document}

\title{ 
\vspace{-2cm}
\begin{flushright}
{\scriptsize CP3-19-20}
\end{flushright}
\vspace{0.5cm}
{\bf \boldmath On the Minimal Mixing of Heavy Neutrinos} \\[8mm]}
\date{}
\author{Marco Drewes \\ \\
{\normalsize \it Centre for Cosmology, Particle Physics and Phenomenology,}\\ {\normalsize \it Universit\'{e} catholique de Louvain, Louvain-la-Neuve B-1348, Belgium}\\
} 
\maketitle
\thispagestyle{empty} 
\begin{abstract}
  \noindent 
We revisit the constraints on the properties of right handed neutrinos from the requirement to explain the observed light neutrino oscillation data in the type-I seesaw model. We use well-known relations to show that there is in general no lower bound on the mixing of a given heavy neutrino with any individual Standard Model  generation. Lower bounds quoted in the literature only apply if the masses of the heavy neutrinos are so degenerate that they cannot be distinguished experimentally. A lower bound on the total mixing (summed over Standard Model generations) can be derived for each heavy neutrino individually, but it strongly depends on the mass of the lightest Standard Model neutrino and on the number of heavy neutrinos that contribute to the seesaw mechanism. Our results have implications for the perspectives of future colliders or fixed target experiments to rule out certain mass ranges for heavy neutrinos.
\end{abstract}

\begin{small}
\tableofcontents
\end{small}
\newpage
\section{Introduction}
\paragraph{Motivation.} Adding right handed neutrinos $\nu_R$ to the Standard Model (SM) of particle physics is probably the most straightforward way to give masses to the light neutrinos that can explain the observed neutrino flavour oscillations. 
To generate a Dirac mass it is sufficient to assume that the right handed neutrinos $\nu_{R i}$ couple to the left handed leptons $\ell_{L a}$ via Yukawa interactions, 
analogously to the way how charged right handed and left handed leptons do.
However, the Yukawa coupling constants  $F_{ai}$
between SM generation $a=e,\mu,\tau$ and heavy neutrino family $i$ would have to be very tiny to explain the smallness of the light neutrino masses ($|F_{ai}|\sim10^{-12}$). 
The appearance of such tiny numbers can be avoided if the $\nu_{R}$ have a Majorana mass $M_M$,
then they can give small Majorana masses $m_i$ to the light neutrinos via the \emph{type I seesaw mechanism}~\cite{Minkowski:1977sc, GellMann:1980vs, Mohapatra:1979ia, Yanagida:1980xy, Schechter:1980gr, Schechter:1981cv}, cf. appendix \ref{App:BasicRelations}.
$F$ and $M_M$ are  $3\times n$ and $n\times n$ matrices, respectively, where $n$ is the number of right handed neutrino flavours $\nu_{Ri}$ added to the SM.
The eigenvalues of $M_M$ define the \emph{seesaw scale(s)}.
The seesaw mechanism in principle works for almost any choice of the seesaw scale. The reason is that the requirement to explain the light neutrino oscillation data only constrains the combination 
\begin{equation}\label{lightmass}
m_\nu = - v^2F M_M^{-1} F^T. 
\end{equation}
Here $v$ is the Higgs field expectation value.
The light neutrino mass squares $m_i^2$ are given by the eigenvalues of $m_\nu^\dagger m_\nu$.
Since $M_M$ and $F$ are both unknown, almost any choice of $M_M$ can be made consistent with the observed data by an appropriate choice of $F$.

The implications of the $\nu_R$'s existence for other observables in particle physics and cosmology strongly depend on the choice of the seesaw scale, cf. e.g. \cite{Drewes:2013gca} for a review. For instance, they can explain the observed matter-antimatter asymmetry in the early universe that is believed to be the origin of baryonic matter in the universe\footnote{See. e.g. ref.~\cite{Canetti:2012zc} for a discussion of the evidence for a matter-antimatter asymmetry in the observable universe and its connection to the origin of matter.} via \emph{leptogenesis} \cite{Fukugita:1986hr}, either during the freeze-out and decay of the heavy neutrinos \cite{Fukugita:1986hr} ("freeze out scenario") or through oscillations during their production \cite{Akhmedov:1998qx,Asaka:2005pn,Hambye:2016sby} ("freeze in scenario"). For sufficiently small Yukawa couplings $F_{ai}$ they can also provide a viable Dark Matter candidate \cite{Dodelson:1993je,Shi:1998km}, cf. \cite{Adhikari:2016bei,Boyarsky:2018tvu} for recent reviews. Finally, if some eigenvalues of $M_M$ are in the (sub) eV range, they may explain the anomalies observed in some neutrino oscillation experiments, cf. \cite{Abazajian:2012ys} for a review.
 
\paragraph{The active-sterile mixing.} A key prediction of the seesaw mechanism is the existence of two sets of neutrino mass eigenstates after electroweak symmetry breaking, consisting of $3$ and $n$ states.
The first set consists of three light neutrinos $\upnu_i$, which can be identified with the usual light neutrino mass eigenstates, while the other $n$ form a set of heavy and almost sterile neutrinos $N_i$.
They can be represented by flavour vectors of Majorana spinors,
\begin{eqnarray}
\upnu \simeq U_{\nu}^{\dagger} \left(\nu_{L} - \theta \nu_{R}^c\right) + \text{c.c.}
\quad , \quad
N \simeq U_N^\dagger \left( \nu_{R} +  \theta^T\nu_{L}^{c}\right) + \text{c.c.}.
\end{eqnarray}
Here $c.c.$ refers to the $c$-conjugation which e.g.~acts as $\nu_R^c=C\overline{\nu_R}^T$ with $C=\ii\gamma_2\gamma_0$.
$U_\nu$ is the 
usual light neutrino mixing matrix and $U_N$ its equivalent amongst the heavy neutrinos. 
We  use the tree level relation and expand to leading order in the mixing between left and right handed neutrinos,  
which is quantified by the matrix
\begin{equation}\label{ThetaDef}
\theta = v F M_M^{-1}.
\end{equation}
Both of theses approximations are justified in the regime of small mixing angles, as discussed in more detail in appendix \ref{Appendix:approximations}.
If some of the eigenvalues of $M_M$ are below the  TeV scale, then the corresponding heavy neutrinos $N_i$ can be produced in experiments. 
The production occurs via the standard weak interaction in the same processes as for ordinary neutrinos, but with amplitudes that are suppressed by the mixing angles $\theta_{ai}$, and with a modified phase space due to the larger heavy neutrino masses $M_i$ \cite{Shrock:1980ct,Shrock:1981wq}.
Hence, assuming that the production is kinematically possible, the production cross section of heavy neutrino flavour $i$ from a decay involving SM flavour $a$ is controlled by the quantities 
\begin{equation}\label{UaiDef}
U_{ai}^2=|\Theta_{ai}|^2 \quad {\rm with} \quad \Theta = \theta U_N^*
\end{equation}
If we start in the flavour basis where $M_M$ is diagonal then we can in general approximate $U_N= 1$ in the following discussion 
 unless the splitting between the eigenvalues of $M_M$ is smaller than the light neutrino masses. We discuss this special case in appendix \ref{Appendix:approximations}. 
It is convenient to also introduce the quantities 
\begin{equation}
U_i^2=\sum_a U_{ai}^2 \quad , \quad U_a^2=\sum_i U_{ai}^2.
\end{equation}
The $U_i^2$ provide a measure for the total interaction strength of a given heavy neutrino $N_i$. 
The $U_a^2$ are a measure of the coupling of all heavy neutrinos to a given SM generation, they are physically most meaningful when all the heavy neutrinos have degenerate masses.

\paragraph{Heavy neutrinos in future experiments.}
Heavy neutrinos can be produced in particle collisions through their $\theta$-suppressed weak interaction. 
For masses $M_i$ above the electroweak scale the sensitivity of LHC experiments \cite{Das:2016hof,Pascoli:2018heg}
can only slightly surpass the range of $U_{ai}^2$ that has already been ruled out indirectly \cite{Atre:2009rg,Kusenko:2009up,Drewes:2013gca,Deppisch:2015qwa,Lindner:2016bgg,Cai:2017mow} if the heavy neutrinos are only produced via their $\theta$-suppressed weak interactions,\footnote{Note that a much better sensitivity can be achieved in extended seesaw models that e.g. include additional gauge interactions; references can be found in the reviews~\cite{Deppisch:2015qwa,Cai:2017mow}, cf. also ref.~\cite{Nemevsek:2018bbt} for a recent study.}
and the lower bounds discussed in the following are practically irrelevant.
The experimental situation is much more promising for $M_i$ below the electroweak scale, where the $N_i$ can be produced copiously in the decays of on-shell  SM particles.
The LHC main detectors are expected to probe mixing angles as small as $U_{\mu i}^2\sim 10^{-8}$ in displaced vertex searches for masses below $20$ GeV \cite{Helo:2013esa,Izaguirre:2015pga,Antusch:2016vyf,Gago:2015vma,Mermod:2017kjb,Antusch:2017hhu,Abada:2018sfh,Cottin:2018nms,Boiarska:2019jcw,Drewes:2019fou,Liu:2019ayx}
and $U_{\mu i}^2\sim10^{-6}$ in prompt decays for masses between $20$ GeV and the $W$ boson mass \cite{Izaguirre:2015pga,Dib:2017iva,Dib:2019ztn}, with an additional order of magnitude gain in sensitivity for the HL-LHC due to the larger integrated luminosity.
Further improvement can be achieved if the muon chambers are used for the reconstruction of displaced vertices\footnote{This idea was originally proposed in the context of searches for supersymmetric particles \cite{Bobrovskyi:2011vx}.} \cite{Boiarska:2019jcw,Drewes:2019fou}.
Future colliders could push the sensitivity  
further down to $U_{\mu i}^2 <  10^{-10}$ \cite{Blondel:2014bra,Antusch:2015mia,Antusch:2017pkq,Antusch:2015rma,Benedikt:2018qee,Antusch:2016ejd}. 
For masses of a few GeV, the $N_i$ can be produced in meson decays \cite{Gorbunov:2007ak,Cvetic:2010rw,Asaka:2016rwd,Bondarenko:2018ptm}.
In this mass range fixed target experiments are generally more sensitive than colliders \cite{Beacham:2019nyx}, though additional dedicated LHC detectors \cite{Chou:2016lxi,Kling:2018wct,Gligorov:2017nwh,Curtin:2018mvb,Dercks:2018wum} could have a similar reach \cite{Alimena:2019zri,Helo:2018qej,Curtin:2018mvb}, and the sensitivity of the main detectors could be increased if data from heavy ion runs is analysed \cite{Drewes:2018xma}.
The NA62 experiment is expected to reach a sensitivity better than $U_{ai}^2\sim 10^{-7}$ between the kaon and D meson mass (in dump mode) \cite{Drewes:2018gkc} and even below $U_{ai}^2\sim 10^{-8}$ below the kaon mass (in kaon mode), 
Similar searches are performed at T2K \cite{Abe:2019kgx} and have been proposed for DUNE \cite{Krasnov:2019kdc,Ballett:2019bgd}.
The SHiP experiment \cite{Alekhin:2015byh,Anelli:2015pba} could push this to $\sim10^{-8}$ below the B meson mass and almost $\sim10^{-10}$ below the D meson mass \cite{SHiP:2018xqw}.
In combination, these searches will be able to cover a significant fraction of the region where the right handed neutrinos alone can simultaneously explain the light neutrino masses and the baryon asymmetry of the universe, cf. e.g. \cite{Chun:2017spz} for a recent review and \cite{Eijima:2018qke,Abada:2018oly} for updated parameter space studies.

\paragraph{A lower bound on the sterile neutrino mixing?} In view of these excellent experimental perspectives it is instructive to ask whether there is a "floor" in the $M_i$-$U_{ai}^2$ planes towards which experiments are moving, i.e., a strict lower bound on $U_{ai}^2$ as a function of $M_i$. 
It is clear that at least some of the $N_i$ must exhibit some mixing with the left handed neutrinos. This becomes evident by re-expressing eq.~\eqref{lightmass} with \eqref{ThetaDef} as
\begin{equation}\label{lightmass2}
m_\nu = -\theta M_M \theta^T.
\end{equation}
Obviously $m_\nu$ in the seesaw relation \eqref{lightmass2} vanishes identically if all $\theta_{ai}$ are zero, therefore the requirement to explain the neutrino masses $m_i$ must impose a lower bound on the $U_{ai}^2$ or combinations of them.
These bounds, sometimes referred to as \emph{seesaw line} in the mass-mixing plane, have been presented in various reviews and are 
frequently quoted in experimental proposals.
This prominence makes it imperative to clearly quote the assumptions that enter the derivations of these lower bounds.
In the present work we use well-known formulae to revise these assumptions, their motivation and their effect for different choices of $n$.

\section{Lower bounds on the sterile neutrino mixing}

\subsection{Absence of a lower bound on $U_{ai}^2$}\label{GeneralConsiderations}
\paragraph{General considerations.}
For illustrative purposes we first consider a hypothetical world with only one SM fermion generation and $n=1$. In this case $m_\nu$, $M_M$ and $\theta$ are numbers (rather than matrices), and the seesaw relation \eqref{lightmass2} uniquely predicts $\theta$ for given values of the masses, 
\begin{equation}\label{NaiveSeesaw}
\theta^2 = m_\nu/M_M.
\end{equation}
In the presence of three SM generations and $n\geq2$ heavy neutrino flavours, knowledge of the $m_i$ and $M_i$ alone is not sufficient any more to predict all entries of the matrix $\theta$. The seesaw relation is also affected by the light neutrino mixing matrix $U_\nu$ as well as other unknown parameters.
If the type-I seesaw is the sole origin of neutrino masses, then one flavour of right handed neutrinos is required for each non-zero light neutrino mass splitting. That is, $n\geq2$ if the lightest SM neutrino is massless ($m_{\rm lightest}=0$) and $n\geq3$ if the lightest SM neutrino is massive ($m_{\rm lightest}>0$).

The connection between light neutrino oscillation data and $\theta$
has been the subject of by various studies, see e.g.~\cite{Casas:2001sr,Gorbunov:2007ak,Kersten:2007vk,Shaposhnikov:2008pf,Gavela:2009cd,Ruchayskiy:2011aa,Asaka:2011pb,Gorbunov:2013dta,Hernandez:2016kel,Drewes:2016jae,Drewes:2018gkc}. 
A particularly convenient way to express it is given by the Casas-Ibarra parameterisation \cite{Casas:2001sr},
\begin{equation}\label{CasasIbarraDef}
\Theta_{ai} = \iu (U_\nu)_{aj} \sqrt{\frac{m_j}{M_i}} \mathcal R_{ji} 
\end{equation}
where $\mathcal{R}$ is an arbitrary matrix that fulfils the condition $\mathcal{R}\mathcal{R}^T=1$.
Eq.~\eqref{CasasIbarraDef} can be obtained from \eqref{lightmass2} and the knowledge that the matrices $U_\nu$ and $U_N$ bring the Majorana mass matrix $m_\nu$ and its equivalent for the heavy neutrinos $M_N$ into a diagonal  forms, 
cf. eq.~\eqref{Diagonalisation}.
Since we work in the flavour basis where the Yukawa couplings of the charged
leptons are diagonal, 
we can identify $U_{\nu}$ with the Pontecorvo-Maki-Nakagawa-Sakata matrix. Here we neglect the deviation of the light neutrino mixing matrix from unitarity caused by the $N_i$.\footnote{
Throughout this paper we work to second order in $\theta$, which in principle yields a light neutrino mixing matrix
$V_\nu = (1 - \frac{1}{2}\theta\theta^\dagger ) U_\nu$.
However, when inserting this into eq.~\eqref{CasasIbarraDef} the approximation $U_\nu=V_\nu$ is consistent with this expansion.}

To see that there is no lower bound on the individual $U_{ai}^2$, one can use \eqref{CasasIbarraDef} to rewrite the requirement $\Theta_{ai}=0$ as a condition on the matrix $\mathcal{R}$,
\begin{eqnarray}\label{Rcondition}
\mathcal{R}_{2i} = - \frac{(U_\nu)_{a1}}{(U_\nu)_{a2}}\sqrt{\frac{m_1}{m_2}} \mathcal{R}_{1i} - \frac{(U_\nu)_{a3}}{(U_\nu)_{a2}}\sqrt{\frac{m_3}{m_2}} \mathcal{R}_{3i}.
\end{eqnarray}
The only other condition on $\mathcal{R}$ is $\mathcal{R}\mathcal{R}^T=1$, which can always be fulfilled simultaneously with \eqref{Rcondition} for at least one choice of $a$ and $i$.\footnote{
To verify this we consider the six independent complex second order polynomials $\delta_{cd}=\sum_{j=1}^n \mathcal{R}_{cj}\mathcal{R}_{dj}$ with $c,d=1,2,3$
obtained from $\mathcal{R}\mathcal{R}^T=1$.
For convenience we choose to impose a condition on the second row of $\mathcal{R}$ because $m_2$ is non-zero for either light neutrino mass hierarchy; other choices are of course equally valid.
The equation with $c=d=2$ always has a solution because $n\geq2$, i.e., 
there is at least one other unconstrained parameter $\mathcal{R}_{2 j}$ with $j\neq i$ in it that can be adjusted.  
Each of the remaining two independent equations 
with either $c=2$ or $d=2$ 
contain at least $n$ unconstrained elements of $\mathcal{R}$ 
(one of which is fixed by each of these equations) because so far no restrictions on the first and third row of $\mathcal{R}$ have been imposed.
Since $n\geq2$  there is at least one independent unconstrained element of $\mathcal{R}$ in each of the three remaining independent equations in which neither $c$ nor $d$ equals $2$.}
This shows that at least one arbitrarily chosen element $\Theta_{ai}$ can be set to zero by appropriate choice of the matrix $\mathcal{R}$, irrespective of the value of $n$ or the mass spectrum of both, the light and the heavy neutrinos.

\paragraph{Application to specific choices of $n$.}
It is instructive to study which elements of the matrix $\Theta$ can be set to zero simultaneously.
The matrix $\mathcal{R}$ contains $3\times n$  complex parameters. 
The condition $\mathcal{R}\mathcal{R}^T=1$ imposes six complex equations on those.
For each element $\Theta_{ai}$ that we demand to vanish the constraint \eqref{Rcondition} adds only one complex equation to this.
This suggests that one can set up to $3\times n - 6$ mixing angles $\Theta_{ai}$ to zero by appropriate choice of $\mathcal{R}$. 

For $n=3$ this equals three, which suggests that one can completely decouple one of the heavy neutrinos ($U_i^2=0$) by setting its mixings with all three SM generations to zero. 
However, this would require that the relation \eqref{Rcondition} is simultaneously fulfilled for $a=e,\mu,\tau$ (with $i$ fixed). The only solution to these three equations is $\mathcal{R}_{1i}=\mathcal{R}_{2i}=\mathcal{R}_{3i}=0$. 
It is straightforward to check that this condition is incompatible with the property $\mathcal{R}\mathcal{R}^T=1$. 
Hence, it is in general not possible to set any $U_i^2$ to zero.
The case when one light neutrino is massless requires special consideration. 
In this case only two rows of $\mathcal{R}$ are physical because one row is multiplied with zero in \eqref{CasasIbarraDef}, and the condition $\mathcal{R}\mathcal{R}^T=1$ should only be applied to the $2\times n$ submatrix of $\mathcal{R}$ that multiplies the two non-zero light neutrino masses. 
This permits the solution $\mathcal{R}_{1i}=\mathcal{R}_{2i}=\mathcal{R}_{3i}=0$ for one of the heavy neutrinos, i.e., one column of $\mathcal{R}$ can be set to zero.
Hence, for $n=3$ one can only fully decouple one of the heavy neutrinos  ($U_i^2=0$)  if $m_{\rm lightest}=0$. 

For similar reasons the simple counting that $3\times n - 6$ mixing angles can be set to zero cannot be applied to the model with $n=2$.
In this case $\mathcal{R}$ only has two rows, and only
four of the entries in $\mathcal{R}\mathcal{R}^T$ are physically meaningful. 
This leads to three conditions from the requirement $\mathcal{R}\mathcal{R}^T=1$. One can set exactly one of the $\Theta_{ai}$ to zero, which uniquely fixes $\mathcal{R}$ in terms of light neutrino parameters.

Finally, let us consider models with $n>3$. Since the model with $n=3$ can already explain the properties of the light neutrinos for any choice of $m_{\rm lightest}$ it is clear that oscillation data cannot impose a lower bound on the couplings of any heavy neutrinos beyond the number of three - even if all of their couplings vanish, the three heavy neutrinos that do mix with the SM neutrinos can explain all experimental data.

The above considerations are consistent with the well-known fact that one sterile neutrino with non-zero mixing is needed for each non-zero light neutrino mass.

\subsection{The model with two sterile families: lower bounds on $U_a^2$, $U_i^2$ and $U_{a}^2/U_b^2$}
The choice $n=2$ is interesting because it is the smallest number that allows to explain the observed neutrino oscillation data (with $m_{\rm lightest}=0$).
It has been stated in several studies (e.g.~\cite{Ruchayskiy:2011aa,Asaka:2011pb,Canetti:2012kh,Drewes:2016jae,Chianese:2018agp}) that the requirement to explain the light neutrino properties in the model with $n=2$ imposes a lower bound on the heavy neutrinos' mixing with individual SM generations. 
This appears to be in contrast to what we found in the previous section \ref{GeneralConsiderations}.
To clarify this point, we  express the $U_{ai}^2$ in terms of the light and heavy neutrino masses, the light neutrino mixing matrix $U_\nu$ and a complex angle $\omega$ that parameterises the matrix $\mathcal{R}$,
\begin{align}\label{Rwithn2}
    \mathcal{R}^\text{NO}
 &= \begin{pmatrix}
        0 & 0
     \\ \cos \omega & \sin \omega
     \\ -\xi \sin \omega & \xi \cos \omega
    \end{pmatrix}
\ ,
 &  \mathcal{R}^\text{IO}
 &= \begin{pmatrix}
        \cos \omega & \sin \omega
     \\ -\xi \sin \omega & \xi \cos \omega
     \\ 0 & 0
    \end{pmatrix}
\ ,
\end{align}
with $\xi = \pm 1$. Here "NO" and "IO" refer to "normal ordering" and "inverted ordering" amongst the light neutrino masses $m_i$.


To be specific, let us consider the mixing between $N_1$ with SM generation $a$,\footnote{Note that choosing $N_1$ is without loss of generality because swapping the signs of $\xi$, $\Im \omega$ and $M_2 - M_1$ while simultaneously changing $\Re \omega\to\pi-\Re \omega$ swaps the labels of $N_1$ and $N_2$, with no physical consequences.} 
with normal ordering of the light neutrino masses. The  case of inverted ordering can be discussed equivalently.
From the condition \eqref{Rcondition} we find a simple condition for $\omega$,
\begin{eqnarray}\label{omegacondition}
\cos \omega = \frac{(U_\nu)_{a3}}{(U_\nu)_{a2}}\sqrt{\frac{m_3}{m_2}}\xi \sin\omega
\end{eqnarray}
This equation always has solutions in the complex $\omega$ plane. If we use it to fix $\omega$ we obtain
\begin{eqnarray}\label{Ua1CanBeMadeZero}
\Theta_{a1} = 0 \quad , \quad \Theta_{a2} = \pm\ii \sqrt{
\frac{m_2}{M_2} (U_\nu)_{a2}^2 + \frac{m_3}{M_2} (U_\nu)_{a3}^2
} 
\end{eqnarray}
Hence, the coupling of a given heavy neutrino $N_1$ to any generation of SM leptons can always be set to zero by choosing suitable values of the experimentally unconstrained parameter $\omega$, 
irrespectively of the heavy neutrino mass spectrum.  Practically this means that there it \emph{no seesaw line in the $M_i$-$U_{ai}^2$ plane}.

The reason why several previous studies claimed that there is a lower bound on the mixing of the heavy neutrinos with individual SM flavours is that they have made the assumption that the heavy neutrino masses  are quasi-degenerate. If the splitting between them is smaller than the resolution of a given direct search experiment, then the resonances cannot be resolved, and the experiment is practically only sensitive to the sum of mixings $U_a^2=U_{a1}^2+U_{a2}^2$.\footnote{
It is worthwhile to note that interference effects due to heavy neutrino oscillations may still allow to recover information about their mass difference even if it cannot be resolved kinematically ~\cite{Boyanovsky:2014una,Anamiati:2016uxp,Dib:2016wge,Das:2017hmg, Antusch:2017ebe, Antusch:2017pkq, Cvetic:2018elt, Hernandez:2018cgc}.
}
From eq.~\eqref{Ua1CanBeMadeZero} it is clear that setting the mixings of one sterile neutrino $N_1$ with a given SM generation to zero necessarily means that $N_2$ must have non-zero mixing with that generation, i.e.,
\emph{there is a seesaw line in the $M_i$-$U_{a}^2$ plane}.

In addition, the requirement \eqref{omegacondition} entirely fixes $\mathcal{R}$, leaving no more freedom to set any other element of $\Theta$ to zero. 
Using the above relations, it is straightforward to check that setting one element $\Theta_{ai}$ to zero necessarily implies that all other entries of $\Theta$ are non-zero if $n=2$.\footnote{To reach this conclusion we have replaced the light neutrino parameters by their current best fit values given in ref.~\cite{Esteban:2018azc}.} 
This in particular implies that there  are lower bounds on both $U_i^2$ for a given choice of the heavy neutrino masses $M_1$ and $M_2$.

Finally, one can show that current light neutrino oscillation data imposes significant constraints on the \emph{flavour mixing pattern} for $n=2$, i.e., on the quantities $U_a^2/(U_e^2+U_\mu^2+U_\tau^2)$ cf. e.g. refs.~\cite{Hernandez:2016kel,Drewes:2016jae}. A detailed discussion of the constraints on the flavour mixing pattern can be found in ref.~\cite{Drewes:2018gkc}.
The quantities $U_a^2$ obviously have little physical meaning if the masses $M_1$ and $M_2$ are sufficiently different that they can be resolved experimentally.

\subsection{The model with three sterile families: lower bounds on $U_i^2$}
The choice $n=3$ is motivated by at least two arguments. From a bottom up viewpoint, it is simply the smallest number of sterile neutrinos that would allow to give mass to all three SM neutrinos, i.e. the minimal choice if $m_{\rm lightest}>0$. From a top down viewpoint, if the seesaw is embedded into a model with an extended gauge group that contains a $U(1)_{B-L}$, then the number of sterile families must equal the number of SM generations to ensure anomaly freedom.
For $n=3$ we can explicitly write
\begin{align}
\mathcal{R} = \mathcal{R}^{23}\mathcal{R}^{13}\mathcal{R}^{12}\;,
\label{Rorder}
\end{align}
where the $\mathcal{R}^{ij}$ have the non-zero elements
\begin{eqnarray}
\mathcal{R}^{ij}_{ii} = \mathcal{R}^{ij}_{jj} = \cos\omega_{ij}, \quad
\mathcal{R}^{ij}_{ij} = -\mathcal{R}^{ij}_{ji} = \sin\omega_{ij}, \quad
\mathcal{R}^{ij}_{kk} = 1; k \neq i,j.
\end{eqnarray}
The matrix  $\mathcal{R}$ can be interpreted as as a complex $3\times 3$ rotation matrix; it can be parameterised by complex "Euler angles" $\omega_{ij}$. 
The condition \eqref{Rcondition} shows that one can always set a specific $U_{ai}^2$ to zero and still explain the observed neutrino oscillation data by choosing appropriate $\omega_{ij}$.
This holds for any light neutrino mass ordering and any value of $m_{\rm lightest}$ and the unknown phases in $U_\nu$. 

On the other hand, there is a lower bound on the sum $U_i^2$. 
This can be seen by using \eqref{CasasIbarraDef} and the unitarity of $U_\nu$ to write 
\begin{eqnarray}\label{LowerUi2}
U_i^2 = \sum_a \sum_{j,k} 
(U_\nu)_{aj} (U_\nu)_{ak}^*
\mathcal{R}_{ji} \mathcal{R}_{ki}^*\frac{\sqrt{m_j m_k}}{M_i}
= \sum_{j,k} 
\mathcal{R}_{ji} \mathcal{R}_{ki}^*\frac{\sqrt{m_j m_k}}{M_i}\delta_{jk}
=\sum_j |\mathcal{R}_{ji}|^2\frac{m_j}{M_i}.
\end{eqnarray}
This is necessarily larger than zero if all $m_j$ are larger than zero because at least one element in the $i$-column of $\mathcal{R}$ is non-zero. 

\begin{figure}
\centering
\begin{tabular}{c c}
normal ordering
&
inverted ordering
\\
\includegraphics[width=0.45\textwidth]{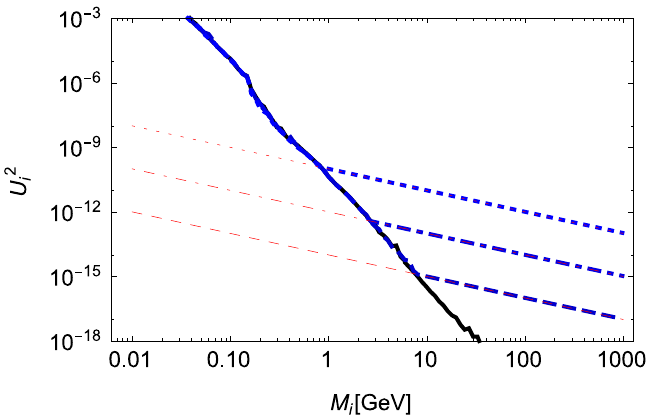}
&
\includegraphics[width=0.45\textwidth]{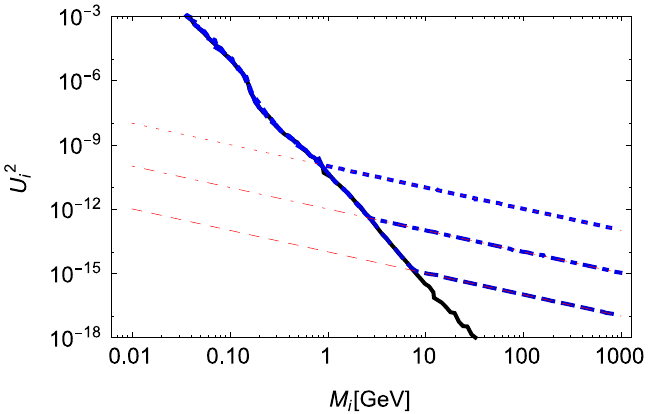}
\end{tabular}
\caption{
The thick lines represent 
a numerical determination of
the lower bounds on the smallest $U_i^2$ 
in the model with $n=3$
from the requirement to simultaneously explain light neutrino oscillation data and respect the constraint on the lifetime from BBN \cite{Ruchayskiy:2012si} 
for 
$m_{\rm lightest}=0 \ {\rm eV}$ (solid),
$m_{\rm lightest}=10^{-5} \ {\rm eV}$ (dashed),
$m_{\rm lightest}=10^{-3} \ {\rm eV}$ (dot-dashed) and
$m_{\rm lightest}=10^{-1} \ {\rm eV}$ (dotted).
The thin red lines compare this to the estimates \eqref{NOestimate} and \eqref{IOestimate}.
As indicated by those estimates there is no difference between the two mass orderings.
We do not show upper bounds from experiments because they usually apply to specific combinations of the $U_{ai}^2$, and a translation into bounds on the total mixings $U_i^2$ relies on computationally expensive numerical studies. Such studies have e.g.~been performed for $n=3$ in refs.~\cite{Fernandez-Martinez:2016lgt} and \cite{Drewes:2015iva} with $M_i$ above and below the electroweak scale, respectively. 
\label{Abbildung}
}
\end{figure}

The precise values of the lower bounds on the $U_i^2$ depend on both, the light and heavy neutrino mass spectrum.
We can estimate them by considering the specific case $\mathcal{R}=1$, where one finds the simple relation
   $\Theta_{ai} = \iu (U_\nu)_{ai} \sqrt{m_i/M_i}$
that reproduces the "naive" seesaw formula \eqref{NaiveSeesaw} if one neglects the effect of light neutrino mixing. 
This implies 
\begin{eqnarray}
&&{\rm normal} \ {\rm ordering}, \ \mathcal{R}=1: \quad \label{NOestimate}\\ 
&&U_1^2 \simeq  \frac{m_{\rm lightest}}{M_1} \ , \ U_2^2 \simeq \frac{\sqrt{m_{\rm lightest}^2+\Delta m_{\rm sol}^2}}{M_2} \ , \ U_3^2 \simeq \frac{\sqrt{m_{\rm lightest}^2+\Delta m_{\rm atm}^2}}{M_3}\nonumber\\
&&{\rm inverted} \ {\rm ordering}, \ \mathcal{R}=1: \quad \label{IOestimate}\\ 
&&U_1^2 \simeq \frac{\sqrt{m_{\rm lightest}^2+\Delta m_{\rm atm}^2
-\Delta m_{\rm sol}^2
}
}{M_1} \ , \ U_2^2 \simeq  \frac{\sqrt{m_{\rm lightest}^2+\Delta m_{\rm atm}^2}}{M_2} \ , \ U_3^2 \simeq  \frac{m_{\rm lightest}}{M_3}.\nonumber
\end{eqnarray}
Here $\Delta m^2_{\rm sol} \simeq 7.4\times10^{-5} {\rm eV}^2$ and $\Delta m^2_{\rm atm} \simeq 2.5\times10^{-3} {\rm eV}^2$ are the two observed neutrino mass splittings.
For $\mathcal{R}\neq1$ the dependence on the various mass scales is more complicated.
If we angles $\omega_{ij}$ are taken to be real, then $\mathcal{R}$ is an actual rotation in the space of sterile flavours. 
Sizeable imaginary parts of the $\omega_{ij}$ give rise to an exponential enhancement of the $\Theta_{ai}$, which can be seen easily when evaluating expressions like $\cos\omega_{ij}$ and $\sin\omega_{ij}$ that appear in $\mathcal{R}$. 
In figure \ref{Abbildung} we verify numerically that the relations \eqref{NOestimate} and \eqref{IOestimate} obtained from setting $\mathcal{R}=1$ provide an excellent estimate for the lower bound on the summed mixing $U_i^2$.
In contrast to that, from \eqref{Rcondition} it is evident that setting $\mathcal{R}=1$ in general does not minimise individual $U_{ai}^2$.
These results are consistent with the numerical studies in refs.~\cite{Gorbunov:2013dta,Drewes:2015iva}, cf. also \cite{Krasnov:2018odt}.

\subsection{Adding the BBN bound: A lower bound on $U_i^2$ for any number of sterile families}

A minimal amount of mixing is required to ensure that heavy neutrinos which populated the early universe decayed before the formation of light elements during \emph{big bang nucleosynthesis} (BBN) in the early universe. Otherwise their decay would affect the abundances of light elements in the intergalactic medium or the effective number of degrees of freedom in the \emph{cosmic microwave background} (CMB) and violate the reasonably good agreement between observation and theory regarding these quantities. 
The resulting constraints on the lifetimes $\tau_i$ of the $N_i$ have most recently been studied in ref.~\cite{Ruchayskiy:2012si}.
The heavy neutrino lifetime roughly scales as $\tau_i\sim U_i^{-2}M_i^{-5}$ \cite{Gorbunov:2007ak}.
It is clear that the requirement to decay before BBN
alone can never impose a lower bound on individual $U_{ai}^2$: A sufficiently short $N_i$ lifetime can e.g. be made consistent with arbitrarily small $U_{\mu i}^2$ if $U_{\tau i}^2$ is large enough to ensure a quick decay into tau final states.
BBN and the CMB only impose a lower bound on the sum $U_i^2=\sum_a U_{ai}^2$, cf.~figure \ref{Abbildung}.

It should be noted that the lifetime constraint only applies under the assumption that the heavy neutrinos come into equilibrium in the early universe. Whether or not this happens in the seesaw model depends on the number of heavy neutrinos $n$ and the mass of the lightest SM neutrino $m_{\rm lightest}$ and has been studied in ref.~\cite{Hernandez:2014fha}.
This loophole makes it possible that sterile neutrinos with sufficiently small $U_i^2$ are viable Dark Matter candidates.
Further, the BBN constraint can be softened if the $N_i$ can decay into new particles that have no or only very feeble interactions with the SM. This would also help to avoid many of the experimental constraints \cite{deGouvea:2015euy}.

\section{Discussion}

\paragraph{Summary of results.}
From the previous considerations it is clear that there is no lower bound on the mixing of a specific heavy neutrino $N_i$ with an individual SM generation $a$ for any choice of $n$ that is consistent with light neutrino oscillation data, i.e.,  there is no \emph{seesaw line} in the $M_i$-$U_{ai}^2$ plane, unless one makes extra assumptions about the model parameters. 
Lower bounds on different combinations of the $U_{ai}^2$ can be imposed if one restricts the parameters $n$, $m_{\rm lightest}$ and the heavy neutrino mass spectrum.\footnote{While $M_i$ and $m_{\rm lightest}$ can in principle be measured, the dependence on $n$ is problematic: Even if two or three heavy neutrinos are discovered one can never be sure that there are no additional $N_i$ that contribute to the seesaw mechanism, but have masses and couplings that are experimentally not accessible. 
The minimal model with $n=2$ requires $m_{\rm lightest}=0$ and predicts various relations between the $U_{ai}^2$ \cite{Hernandez:2016kel,Drewes:2016jae,Drewes:2018gkc}.
It is at least in principle fully testable if, in addition to all $M_i$ and $U_{ai}^2$, the Dirac phase $\delta$ in $U_\nu$ and indirect signatures like neutrinoless double $\beta$ decay or flavour observables are measured \cite{Hernandez:2016kel,Drewes:2016jae}. 
If all these observables could consistently be described within the model, this would be a strong hint that the $n=2$ model is realised in nature. For $n>2$ the parameter space is much larger, and it is not clear that enough observables can be found to overconstrain the model.} 
One can therefore only make statements about lower bounds within a given particle physics model in which the seesaw is implemented.\footnote{In this context it is worthwhile to emphasise that the bounds quoted here apply to all $N_i$ that contribute to the generation of light neutrino masses in a measurable way. 
They can also be applied to models that in principle contain a larger number of heavy neutrinos, but only $n$ of them give a measurable contribution to the $m_i$ (e.g. because the mixing angles of the others are much smaller than the estimate \eqref{NaiveSeesaw}). For instance, all bounds that apply to the $n=2$ model also apply to the two heavier heavy neutrinos $N_2$ and $N_3$ in the Neutrino Minimal Standard Model ($\nu$MSM) \cite{Asaka:2005an,Asaka:2005pn} because the couplings of $N_1$ are extremely feeble, see below.}
In summary there are the following lower bounds.
\begin{itemize}
\item[1)] In the model with $n=2$ there  are lower bounds from neutrino oscillation data on the $U_i^2$ \cite{Asaka:2011pb}. 
\item[2)]  In the model with $n=3$ there are lower bounds from neutrino oscillation data on the $U_i^2$  that depend on $m_{\rm lightest}$, cf. eqns.~\eqref{NOestimate} and \eqref{IOestimate}. The lower bound on the smallest of the $U_i^2$ vanishes if $m_{\rm lightest}=0$, so that there is no seesaw line in the $M_i$-$U_i^2$ plane unless one requires all light neutrinos to be massive \cite{Gorbunov:2013dta,Drewes:2015iva}.
\item[3)]  BBN imposes a lower bound on $U_i^2$ for each heavy neutrino $N_i$ that comes into thermal equilibrium in the early universe \cite{Dolgov:2000pj,Dolgov:2000jw,Ruchayskiy:2012si}.  This bound does not directly depend on the light neutrino mass spectrum (in particular $m_{\rm lightest}$), and it applies to each $N_i$ individually (and therefore does not directly depend on $n$). 
However, the lower bounds 1) and 2) imply that both heavy neutrinos must come into thermal equilibrium for $n=2$; for $n=3$ one of the $N_i$ can avoid equilibration if $m_{\rm lightest}<10^{-3} {\rm eV}$ \cite{Hernandez:2014fha}.
\item[4)] In the model with $n=2$ there are lower bounds from neutrino oscillation data on the $U_a^2=\sum_i U_{ai}^2$ \cite{Asaka:2011pb,Ruchayskiy:2011aa}. These quantities are phenomenologically important if the two heavy neutrino masses are degenerate. 
\item[5)] Both, lower and upper bounds on the ratios $U_a^2/(U_e^2+U_\mu^2+U_\tau^2)$ from neutrino oscillation data exist in the model with $n=2$ \cite{Hernandez:2016kel,Drewes:2016jae,Drewes:2018gkc}. 
\item[6)] The combination of the bounds 3), 4) and 5) implies a lower bound from BBN on the $U_a^2$ in the model with $n=2$. More precisely, the combined bounds from BBN and neutrino oscillation data impose a stronger lower bound on the $U_i^2$ \cite{Canetti:2012kh} and the  $U_a^2$ \cite{Drewes:2016jae} than oscillation data alone if the heavy neutrino mass is below a GeV.
\end{itemize}
A key observation in the above list is that lower bounds on the mixings  of heavy neutrinos with individual SM generations only apply if the heavy neutrino masses are quasi-degenerate, cf. points 4)-6). This immediately leads to the question whether there are any reasons why one should assume a degenerate heavy neutrino mass spectrum.

\paragraph{Symmetry considerations.}

The smallness of the light neutrino masses $m_i$ can be explained by utilising the seesaw relation~\eqref{lightmass2} in different ways.
One possibility is that the $N_i$ are superheavy; in this case the smallness of the $m_i$ is due to the smallness of $v/M_i$, where we take the Higgs expectation value $v$ as an order parameter for the electroweak scale.
This conventional seesaw mechanism cannot work for the mass range $M_i<v$ where experiments are most sensitive.
One possibility to achieve small neutrino masses with $M_i<v$ is to simply choose very small Yukawa couplings $|F_{ai}|\simeq \sqrt{m_i M_i }/v \sim 10^{-6} \sqrt{M_i/v}$; 
in this case one would roughly expect mixings of the order \eqref{NaiveSeesaw}.
From an experimental viewpoint there is nothing wrong with this choice, but many theorists consider the idea of coupling constants of this size to be "unnatural".

One possibility to avoid the appearance of small numbers is that the smallness of the $m_i$ is the result of a slightly broken symmetry.
Specific examples that motivate this limit include in \emph{inverse seesaw} type scenarios~\cite{Wyler:1982dd,Mohapatra:1986aw,Mohapatra:1986bd,Bernabeu:1987gr}, a \emph{linear seesaw}~\cite{Akhmedov:1995ip,Akhmedov:1995vm}, scale invariant models~\cite{Khoze:2013oga}, some technicolour-type models~\cite{Appelquist:2002me,Appelquist:2003uu} or the $\nu$MSM~\cite{Shaposhnikov:2006nn}.
This symmetry can e.g. be related to a generalised lepton number $\bar{L}$ \cite{Shaposhnikov:2006nn,Kersten:2007vk} that includes the SM lepton number $L$ and contributions from the $\nu_R$.
For $n=3$ this can be seen explicitly when using the parameterisation \cite{Abada:2018oly}
\begin{equation}
M_M=\M\begin{pmatrix} \Mu'  & 0 & 0\\ 0 & 1 - \Mu  &0  \\ 0 & 0 & 1 + \Mu  \end{pmatrix} \quad , \quad
F=\begin{pmatrix}  F_e\epsilon'_e & F_e(1 + \epsilon_e) & \ii F_e (1 - \epsilon_e)  \\ F_\mu \epsilon'_\mu & F_\mu(1 + \epsilon_\mu) & \ii F_\mu(1 - \epsilon_\mu)  \\ 
 F_\tau\epsilon'_\tau& F_\tau(1 + \epsilon_\tau) & \ii F_\tau(1 - \epsilon_\tau)  \end{pmatrix}\label{FullNeutrinoMass}.
\end{equation}
When expressing $M_M$ and $F$ in this form, there is only one new mass scale, \emph{the seesaw scale}  $\M$. 
A low value of $\M$ can be motivated by technical naturalness because it avoids a contribution of the $N_i$ to the weak hierarchy problem, 
and $B-L$ is conserved for $\M\to0$., where $B$ is the baryon number.
The numbers $F_a$ quantify the overall magnitude of the heavy neutrinos to individual SM flavours; they can in principle be of order unity, so that no small numbers have to be dialed by hand. The numbers $\epsilon_a, \epsilon'_a, \Mu, \Mu'$ represent symmetry breaking parameters.
The combination $B-\bar{L}$ is conserved in the limits 
$\epsilon_a, \epsilon'_a, \Mu \to 0$
or
$\epsilon_a, \Mu, \Mu' \to 0$
which are ofter referred to as \emph{approximate lepton number conservation} or \emph{symmetry protected regime}.
A detailed discussion of the quantity $\bar{L}$ can e.g. be found in  \cite{Abada:2018oly}, cf. also \cite{Moffat:2017feq}.

An important prediction of the $B-\bar{L}$ conserving limit of eq.~\eqref{FullNeutrinoMass} is that the masses of two of the heavy neutrinos become degenerate ($\mu\to0$), while the the third one decouples from the SM ($\epsilon'_a\to0$).
More generally, for any $n$, an approximate $B - \bar{L}$ symmetry can be realised 
if the $\nu_{Ri}$ either come in pairs with equal mass $M_i = M_j$ and couplings $F_{ai} = \ii F_{a j}$ that can be represented by Dirac spinors $\frac{1}{\sqrt{2}}(\nu_{Ri} + \nu_{R i}^c) + \frac{\ii}{\sqrt{2}} (\nu_{Rj} + \nu_{R j}^c)$
or have negligible couplings to the SM, which allows for odd numbers of $n$.
This provides a theoretical motivation for a degenerate heavy neutrino mass spectrum. 
From an experimental viewpoint such scenarios are attractive because the quantities $U_{ai}^2 \sim F_a (v/M_i)$ can be large enough to give sizeable event rates at colliders or fixed target experiments without fine tuning.
This is the reason why many phenomenological studies have focussed on the case of quasi-degenerate heavy neutrinos and considered the quantities $U_a^2$ instead of $U_{ai}^2$.

However, it is important to note that the starting point to motivate the $B-\bar{L}$ symmetric scenario was the need to avoid small Yukawa couplings, and to make sizeable Yukawa couplings technically natural.  
In the present work we are interested in lower bounds on the $U_{ai}^2$, and  the smallest mixing angles necessarily involve small Yukawa couplings (recall that $\theta_{ai}=vF_{ai}/M_i$).
For small Yukawa couplings the technical naturalness is automatically guaranteed by the $B-L$ conservation in the limit $F_{ai}\to 0$, which does not require any specific properties of the heavy neutrino mass spectrum.
Therefore, when it comes to the smallest mixing angles, there is no good reason to assume a mass degeneracy from a  theoretical viewpoint.

\paragraph{Leptogenesis.}
Low scale leptogenesis via the freeze-out mechanism is known to be feasible only if the heavy neutrino masses are quasi-degenerate ("resonant leptogenesis") \cite{Pilaftsis:2003gt}. 
This is not the case for the freeze-in mechanism \cite{Akhmedov:1998qx,Asaka:2005pn}, where the BAU can be reproduced without a heavy neutrino mass degeneracy \cite{Drewes:2012ma,Canetti:2014dka,Abada:2018oly}. 
 An important exception is the model with $n=2$ \cite{Asaka:2005pn}, where a mass degeneracy at the percent level is needed  for leptogenesis \cite{Canetti:2010aw,Canetti:2012vf,Canetti:2012kh} (cf. \cite{Eijima:2018qke,Antusch:2017pkq,Hernandez:2016kel} for recent results). 
 This limitation practically also applies to the $\nu$MSM because one of the heavy neutrinos is so feebly coupled that it can practically be neglected during leptogenesis, see below.
Hence, the requirement for successful leptogenesis motivates a degenerate heavy neutrino mass spectrum only in the specific cases of $n=2$ and in the $\nu$MSM.

\paragraph{The $\nu$MSM.}
The $\nu$MSM \cite{Asaka:2005an,Asaka:2005pn} is a minimal extension of the SM by three right handed neutrinos that aims to explain all observed phenomena which require New Physics \cite{Shaposhnikov:2008pf,Boyarsky:2009ix}\footnote{
Explaining all cosmological data without introducing new particles other than the $\nu_R$ requires the Higgs field to drive cosmic inflation \cite{Bezrukov:2007ep,Bezrukov:2008ut}.
} and could in principle be a complete description of Nature for energies up to the Planck scale \cite{Shaposhnikov:2007nj}. One of the heavy neutrinos ($N_1$) has a mass in the keV range and very feeble couplings; this particle is a warm Dark Matter candidate. 
The other two ($N_2$ and $N_3$) have quasi-degenerate 
masses in the GeV range; they generate the observed light neutrino mass splittings and the baryon asymmetry of the universe. 
For the purpose of the present discussion the $\nu$MSM represents a  specific parameter choice within the seesaw model with $n=3$.
In the parameterisation \eqref{CasasIbarraDef} with \eqref{Rorder} this corresponds to  $M_1 \ll M_2, M_3$ and $(\omega_{12},\omega_{13},\omega_{23})=(0,0,\omega)$. 
In the parameterisation \eqref{FullNeutrinoMass} it corresponds to $\epsilon_a, \epsilon'_a, \Mu, \Mu'\to 0$.

If $N_1$ alone is required to explain the entire Dark Matter density, then the observational constraints on $U_1^2$ \cite{Adhikari:2016bei,Boyarsky:2018tvu} are so strong that its effect on the light neutrino mass generation can be neglected, i.e., $m_{\rm lighest}^2 \ll \Delta m^2_{\rm sol}$ \cite{Boyarsky:2006jm}.
For the discussion of the seesaw mechanism 
and leptogenesis in the $\nu$MSM we can set $m_{\rm lightest}=0$ for all practical purposes, so that $N_1$ completely decouples.
This means that all constraints on the $U_{ai}^2$ and combinations of them that we found for the case $n=2$ also apply to the two heavier $N_i$ in the $\nu$MSM.
One could say that the freedom that \eqref{LowerUi2} offers to push the lower bound on the smallest $U_i^2$ to very small values by playing with $m_{\rm lightest}$ has been "used up" to make the Dark Matter candidate stable.

\section{Conclusions}
We have investigated the possibility to derive a lower bound of the mixing of sterile neutrinos in the type-I seesaw model from the requirement to explain light neutrino oscillation data. We find that there is in general no lower bound on the mixing  $U_{ai}^2$ of a specific heavy neutrino $N_i$ with a given SM generation $a$. 
This conclusion remains unchanged if one takes the cosmological constraints from primordial nucleosynthesis and the CMB into consideration, and even if one adds the additional requirement to explain the baryon asymmetry of the universe via low scale leptogenesis.
In contrast to that, depending on the mass of the lightest neutrino, there can be lower bounds on the sums $U_i^2=\sum_a U_{ai}^2$.
An excellent estimate of those can be obtained by setting the matrix $\mathcal{R}$ in the Casas-Ibarra formula \eqref{CasasIbarraDef} to unity, cf.~eqns.~\eqref{NOestimate} and \eqref{IOestimate} and figure \ref{Abbildung}.

An important exception is the $\nu$MSM, where the requirement to simultaneously explain the observed neutrino oscillation data and the baryon asymmetry with only two heavy neutrinos enforces a mass degeneracy amongst them. This means that experiments cannot resolve the signatures from the two heavy neutrinos individually and are only sensitive to the sum of their mixings $U_a^2=U_{a1}^2+U_{a2}^2$. 
In this situation, lower bounds on $U_a^2$ can be derived that practically act as a "floor" for experimental searches. 
If more than two heavy neutrinos contribute to the generation of the light neutrino masses these lower bounds can be avoided.

It goes without saying that any bound derived from neutrino oscillation data can be loosened if the mixing with right handed neutrinos is not the sole origin of the light neutrino masses. Moreover, if the heavy neutrinos are part of a extended Dark Sector, decays into this sector could also soften the BBN bound.

\section*{Acknowledgements}
I would like to thank Juraj Klaric and Valerie Domcke for proofreading the draft of this manuscript.
I am also grateful to the Max Planck Institute for Physics (Werner Heisenberg Institut) in Munich for their hospitality during the work on this project.

\begin{appendix}

\section{The seesaw model}\label{App:BasicRelations}
The seesaw mechanism has been reviewed numerous times in the literature. Here we provide a derivation of the relevant formulae in order to define our conventions, which are based on the notation used in ref.~\cite{Drewes:2013gca}.


The most general renormalisable extension of the SM that only contains SM fields and $n$ flavours of $\nu_R$ reads
\begin{equation}
    \mathcal L_{\nu_R}
  = \iu \overline{\nu_R}_i \slashed\partial \nu_{Ri}
  - \frac{1}{2} \left( \overline{\nu_R^c}_i(M_M)_{ij}\nu_{Rj}
    + \overline{\nu_R}_i(M_M^\dagger)_{ij}\nu_{R}^c \right)
  - F_{ai} \overline{\ell_{L a}} \varepsilon \phi^* \nu_{Ri}
  - F_{ai}^* \overline{\nu_R}_i \phi^T \varepsilon^\dagger \ell_{L a}
\ . \label{eq:Lagrangian}
\end{equation}
Here SU(2) indices have been suppressed; $\varepsilon$ denotes the totally antisymmetric SU(2) tensor.
The $F_{a i}$ are the Yukawa couplings between the sterile flavours $\nu_{R i}$ and the SM lepton generations $\ell_{L a}$, and $M_M$ is a Majorana mass matrix for the singlet fields $\nu_{R i}$.\footnote{We use four-component spinor notation, but the chiral spinors $\nu_R$ and $\ell_L$ have only two non-zero components ($P_R\nu_R=\nu_R$ and $P_L\ell_L=\ell_L$). Hence, no explicit chiral projectors are required in the weak interaction term \eqref{WeakWW}.}
The spontaneous breaking of the electroweak symmetry by the Higgs expectation value $\phi=(0, v)^T$ (with $v=174$ GeV) generates a Dirac mass term $\overline{\nu_{L}} m_D\nu_{R}$ with $m_D=vF$ from the Yukawa interaction term {$F\overline{\ell_L}\varepsilon\phi^*\nu_{ R }$}. Then the complete neutrino mass term reads
\begin{equation}
\frac{1}{2}(\bar{\nu_L} \ \bar{\nu_R^c})
\underbrace{\begin{pmatrix}{\delta}m_{\nu}^{1loop} & m_D \\ m_D^T & M_M + {\delta}M_{N}^{1loop}\end{pmatrix}}_{ \equiv \, {\mathcal M}}
\begin{pmatrix}
\nu_L^c
\\
\nu_R
\end{pmatrix}\ .
\label{eq:auxlabel}
\end{equation}
Here we have included the 1-loop corrections \cite{Pilaftsis:1991ug}
  in order to be consistent at second order in the Yukawa couplings $F$. 
The full $(3+n)\times(3+n)$ neutrino mass matrix (\ref{eq:auxlabel}) can be diagonalised as
\begin{equation}
\mathcal{U}^{\dagger}\mathcal{M}\mathcal{U}^{\ast}=\begin{pmatrix}m_{\nu}^{\rm diag} & \\ & M_N^{\rm diag} \end{pmatrix}\ .
\label{eq:Mdiagonalization}
\end{equation}
Here $m_{\nu}^{\rm diag}$ and $M_N^{\rm diag}$ are diagonal $3\times3$ and $n\times n$ matrices, respectively.
It is convenient to parametrise the mixing matrix $\mathcal{U}$ as \cite{Fernandez-Martinez:2015hxa}
\begin{equation}
\mathcal{U}= \begin{pmatrix} \cos(\theta) & \sin(\theta) \\ -\sin(\theta^\dagger) & \cos(\theta^\dagger)  \end{pmatrix}
\begin{pmatrix} U_{\nu} & \\ & U_N^{\ast} \end{pmatrix}\ ,
\end{equation}
with the definitions
\begin{align}
\cos(\theta)=\sum_{n=0}^\infty \frac{(-\theta\theta^\dagger)^n}{(2n)!} \quad , \quad
 \sin(\theta)=\sum_{n=0}^\infty \frac{(-\theta\theta^\dagger)^n\theta}{(2n+1)!}.
 \end{align} 
 In the parameterisation (\ref{eq:Mdiagonalization}) the full mass matrix $\mathcal{M}$ is first block-diagonalised by a complex "rotation" that is characterised by the $3 \times n$ matrix of mixing angles $\theta$, and then the unitary matrices $U_\nu$ and $U_N^*$ diagonalise the $3\times3$ and $n\times n$ blocks $m_\nu$ and $M_N$ in the upper left and lower right corners, respectively. More precisely,
 \begin{eqnarray}\label{Diagonalisation}
 U_\nu^\dagger m_\nu U_\nu^* = {\rm diag}(m_1,m_2,m_3)\equiv m_\nu^{\rm diag} \  , \ U_N^T M_N U_N = {\rm diag}(M_1,M_2,\ldots,M_n)\equiv M_N^{\rm diag}.
 \end{eqnarray}
In the \emph{seesaw limit} $|\theta_{a i}|\ll 1$, one can expand to second order in $\theta$. This yields
\begin{eqnarray}\label{eq:Uexpand}
\theta \simeq m_D M_M^{-1} &=& v F M_M^{-1}\,,\quad  \quad \cos(\theta) = 1-\frac{1}{2}\theta\theta^{\dagger} + {\cal O}(\theta^4) \,, \quad  \quad   \sin(\theta) = \theta  + {\cal O}(\theta^3)
\end{eqnarray}
and
\begin{eqnarray}\label{eq:blocks_mass_matrix}
    m_\nu
  =   m_\nu^{\rm tree} + \delta m_\nu^{\rm 1loop} \ &,& \ 
m_\nu^{\rm tree} = - \theta M_M \theta^T\\
M_N = M_N^{\rm tree} + \delta M_N^{\rm 1loop} \ &,& \
M_N^{\rm tree}  = M_M + \frac{1}{2} (\theta^\dagger \theta M_M + M_M^T \theta^T \theta^{*})\label{MN_Def}.
\end{eqnarray}
In the seesaw limit, spectrum of neutrino mass states is separated into two sets, consisting of three light and $n$ heavy mass eigenstates, respectively.  These can be expressed in terms of the Majorana spinors
\begin{align}
    \upnu_i &
  = \left[
  V_{\nu}^{\dagger}\nu_L-U_{\nu}^{\dagger}\theta \nu_R^c+V_{\nu}^T\nu_L^c-U_{\nu}^T\theta^{\ast} \nu_R
  \right]_i
\ ,
  & N_i &
  = \left[
  V_N^\dagger\nu_R+\Theta^T \nu_L^c + V_N^T\nu_R^c+\Theta^{\dagger}\nu_L
  \right]_i
\ . \label{HeavyMassEigenstates}
\end{align}
Here $V_\nu = (1 - \frac{1}{2}\theta\theta^\dagger ) U_\nu$,  $V_N = (1 - \frac{1}{2} \theta^T \theta^*) U_N$.
$V_\nu$ is the usual light neutrino mixing matrix, which exhibits a non-unitarity at second order in $\theta$ due to the existence of the $N_i$.
The active-sterile mixing matrix
 \begin{equation}
\Theta = \theta U_N^*
\end{equation}
quantifies the interactions of the $N_i$ as 
\begin{align}
    \mathcal L
  \supset&
  - \frac{g}{\sqrt{2}}\overline{N}_i \Theta^\dagger_{i a}\gamma^\mu e_{L a} W^+_\mu
 - \frac{g}{2\cos\theta_W}\overline{N_i} \Theta^\dagger_{i a}\gamma^\mu \nu_{L a} Z_\mu
 - \frac{g}{\sqrt{2}}\frac{M_i}{m_W}\Theta_{a i} h \overline{\nu_{L a}}N_i
+ \text{h.c.}
\ . \label{WeakWW}
\end{align}
The first two terms represent the $\theta$-suppressed weak interactions of the $N_i$, while the third term represents the Yukawa coupling to the physical Higgs field $h$ in the unitary gauge.
Here we have employed the relation $m_W = \frac{1}{2} g v$ involving the weak gauge coupling constant $g$.
Through the mixing $\Theta$ the heavy neutrinos $N_i$ can replace ordinary neutrinos in all processes if this is kinematically allowed, but with amplitudes suppressed by the angles~$\Theta_{a i}$.
It is convenient to introduce the quantities \eqref{UaiDef}, which practically govern the event rates for processes involving the $N_i$.

\section{Important approximations and simplifications}\label{Appendix:approximations}
For the analysis presented in the main text of this work we make two important approximations to the formulae presented in sec.~\ref{App:BasicRelations}:
we approximate $U_N\simeq 1$ and
we neglect the loop corrections $\delta m_\nu^{\rm 1loop}$ and $\delta M_N^{\rm 1loop}$.

\subsection{The role of the matrix $U_N$}
The difference between the Majorana mass matrix $M_M$ in the Lagrangian \eqref{eq:Lagrangian} and the physical heavy neutrino mass matrix after electroweak symmetry breaking $M_N$ is of order of the light neutrino masses $m_i$ \cite{Shaposhnikov:2008pf}, as one can see by comparing eqns.~\eqref{eq:blocks_mass_matrix} and \eqref{lightmass2}.
It is therefore in general negligible and we one can approximate $U_N\simeq 1$ unless splitting between the eigenvalues of $M_M$ is of the order of the light neutrino masses $m_i$ (or smaller). Such a degenerate spectrum would appear highly tuned unless it can be explained by a symmetry. 
In eq.~\eqref{FullNeutrinoMass} two of the heavy neutrinos have degenerate masses in the $B-\bar{L}$ symmetric limit, and the mixing amongst those two by $U_N$ can in principle be sizeable. The mixing with the third state is parametrically small because either $\mu'\ll 1$ or $\epsilon'_a\ll1$ in the approximately $B-\bar{L}$ symmetric limit. The implies
\begin{equation}
U_N\simeq
\begin{pmatrix}  
1 & 0 & 0  \\
0 & a &  b  \\ 
0 & -e^{\ii \varphi} b^* &  e^{\ii \varphi} a^*  
\end{pmatrix}.
\end{equation}
Here $a$, $b$ are  complex numbers with $|a|^2+|b|^2=1$ and $\varphi$ is a phase; they parameterise the (approximately) unitary sub-matrix that mixes the degenerate states.
The requirement ${\rm det}U_N=1$ fixes $\varphi=0$.
In the symmetric limit the matrix $M_N^\dagger M_N$, the eigenvectors of which form $U_N$, is real,\footnote{
Using $\theta\simeq F \frac{v}{\M}$ and setting all $B-\bar{L}$ violating parameters in \eqref{FullNeutrinoMass} to zero
it is straightforward to derive from eq.~\eqref{MN_Def} that the tree level contribution 
$M_N^{\rm tree}$
to $M_N$ is real in the $B-\bar{L}$ symmetric limit. 
The loop correction $\delta M_N^{\rm 1loop}$ can be calculated from the Lagrangian \eqref{WeakWW} with $M_N^{\rm tree}$ in the heavy neutrino propagator. 
Since the weak interaction is fundamentally flavour blind, the flavour structure of $\delta M_N^{\rm 1loop}$ can only come from $F$ and $M_M$.  
In the symmetric limit the submatrix of $M_N^{\rm tree}$ for the two heavy neutrinos with non-zero Yukawa couplings is proportional to a unit matrix. Hence, $\delta M_N^{\rm 1loop}$ must have the same overall flavour structure as the tree level term and is also real.
} 
and therefore $U_N$ is also real.
Therefore $a$ and $b$ are can taken to be real.
Now using $\theta\simeq F \frac{v}{\M}$ we find
\begin{equation}
\Theta =\theta U_N^* \simeq \frac{v}{\M}
\begin{pmatrix}  
0 & F_e (a + \ii b) & F_e (\ii a +  b)  \\ 
0 & F_\mu (a + \ii b) & F_\mu (\ii a +  b)  \\
0 & F_\tau ( a + \ii b) & F_\tau (\ii a +  b)  
\end{pmatrix},
\end{equation}
Since $a$ and $b$ are real with $a^2+b^2=1$ we find that 
$|\Theta_{ai}|^2=|\theta_{ai}|^2$ in this limit, and hence  $U_N$ does not affect the $U_{ai}^2$ even if its individual entries are large.
This is consistent with the numerical results found in ref.~\cite{Chianese:2018agp}.

\subsection{Radiative corrections}
The loop corrections $\delta m_\nu^{\rm 1loop}$ and $\delta M_N^{\rm 1loop}$ have two effects. 
First, $\delta m_\nu^{\rm 1loop}$ modifies the relation between $\Theta$ and the observed light neutrino parameters. 
This effect can be taken account of by replacing $M_N^\text{diag}$ in the parameterisation \eqref{CasasIbarraDef} by $\tilde{M}\simeq M_N^\text{diag}[
1 - \frac{M_N^\text{diag}}{v^2} l(M_N^\text{diag}) ]$
with $l(M_i) = \frac{1}{(4\pi)^2}\left[\frac{3\text{ln}[(M_i/m_Z)^2]}{(M_i/m_Z)^2 -  1} + \frac{\text{ln}[(M_i/m_H)^2]}{(M_i/m_H)^2 - 1}\right]$ \cite{Lopez-Pavon:2015cga}, where $m_Z$ and $m_H$ are the Z boson and Higgs boson masses, respectively.
Practically this amounts to multiplying $\Theta$ by a diagonal matrix from the left. This cannot qualitatively change any of our conclusions about the $U_{ai}^2$. 
Second, $\delta M_N^{\rm 1loop}$ modifies the matrix $U_N$ that diagonalises $M_N$. But as shown in the previous paragraph, $U_N$ does not affect the $U_{ai}^2$. It is therefore justified to neglect radiative corrections in the main discussion.

\end{appendix}

\bibliographystyle{JHEP}
\bibliography{ActiveSterileMixing.bib}{}
\end{document}